\documentclass{PoS}
\usepackage{amsmath}
\usepackage{booktabs}
\usepackage[font=small,labelfont=bf]{caption} 
\usepackage[svgnames]{xcolor} 

\usepackage{cancel}
\newcommand\ccancel[2][black]{\renewcommand\CancelColor{\color{#1}}\cancel{#2}}

\usepackage{tikz}
\usepackage{pgfplots}
\usetikzlibrary{arrows.meta} 
\usetikzlibrary{patterns,snakes}

\graphicspath{{figs/}}
\renewcommand{\O}{\mathrm{O}}

\newcommand{\zf}{z_f}
\newcommand{\mcr}{m_{\rm cr}}
\newcommand{\psibar}{\bar{\psi}}

\def\fv{f_{\rm V}}
\def\fa{f_{\rm A}}

\def\gv{g_{\rm V}}

\def\ga{g_{\rm A}}

\def\gx{g_{\rm X}}

\def\ka{k_{\rm A}}
\def\kv{k_{\rm V}}

\def\la{l_{\rm A}}
\def\lv{l_{\rm V}}

\def\ly{l_{\rm Y}}

\title{$\chi$SF near the electroweak scale}

\ShortTitle{$\chi$SF near the electroweak scale}

\author{Isabel Campos \\
Instituto de F\'isica de Cantabria - IFCA-CSIC, Avda. de Los
Castros s/n, 39005 Santander, Spain
}

\author{Mattia Dalla Brida \\
Universit\`a di Milano-Bicocca, Dipartimento di Fisica,  and INFN,
sezione di Milano-Bicocca, Piazza della Scienza 3, I-20126 Milano, Italy
}

\author{Giulia Maria de Divitiis \\
Universit\`{a} di Roma ``Tor Vergata'', Dipartimento di Fisica and INFN sezione
di Roma \mbox{Tor Vergata}, Via della Ricerca Scientifica~1, I-00133 Rome, Italy
}
\author{\speaker{Andrew Lytle} \\
        INFN, Sezione di Roma Tor Vergata, Via della Ricerca Scientifica 1,
        00133 Roma RM, Italy\\
        E-mail: \email{andrew.lytle@roma2.infn.it}}

\author{Mauro Papinutto \\
Universit\`{a} di Roma ``La Sapienza'', Dipartimento di Fisica and INFN sezione
di Roma1, Piazzale Aldo Moro 2, I-00185 Roma, Italy
}

\author{Anastassios Vladikas \\
INFN, sezione di Roma Tor Vergata, c/o Dipartimento di Fisica, 
Via della Ricerca Scientifica~1, I-00133 Rome, Italy
}

\abstract{
We employ the chirally rotated Schr\"odinger functional ($\chi$SF) to study
two-point fermion bilinear correlation functions used in the determination of
$Z_{A,V,S,P,T}$
 on a series of well-tuned ensembles.
 The gauge configurations,  which span renormalisation scales from 4 to 70~GeV, 
 are generated with  $N_{\rm f}=3$ massless flavors and  
Schr\"odinger Functional (SF) boundary conditions. 
Valence quarks are computed with $\chi$SF boundary conditions.
We show preliminary results on the tuning of the $\chi$SF Symanzik coefficient $z_f$ and the scaling of 
the axial current normalization $Z_{\rm A}$. Moreover 
we carry out a detailed comparison with the expectations from one-loop
perturbation theory. Finally we outline how automatically $\O(a)$-improved $B_{\rm K}$ matrix elements, including BSM contributions, 
can be computed in a $\chi$SF renormalization scheme.
}

\FullConference{
    The 37th Annual International Symposium 
    on Lattice Field Theory - LATTICE2019\\
	16-22 June, 2019\\
	Wuhan, China.}

\begin{document}

The chirally rotated Schr\"odinger functional ($\chi$SF) with massless Wilson
fermions is a lattice regularization which endows the Schr\"odinger functional
(SF) with the property of automatic \mbox{$\O(a)$-improvement}.
The $\chi$SF framework is effective in reducing lattice artefacts in
correlation and step scaling functions, but especially it offers
new strategies to study and simplify the pattern of renormalization.
The price to pay for the automatic $\O(a)$-improvement is the nonperturbative
tuning of coefficients of new boundary counterterms.
This tuning is the first phase of a long-term project,
aiming at the computation of $B_{\rm K}$ low-energy contributions beyond the Standard Model (BSM),
with Wilson fermion $N_{\rm f} = 2+1$ lattice QCD in a non-unitary
(mixed-action) framework.

\section{The $\chi$SF setup}
\noindent Following ref.~\cite{Sint:2010eh}, the fermion flavour doublet $\psi = \begin{pmatrix}  \psi_u \\ \psi_d
\end{pmatrix}$ satisfies  $\chi$SF boundary conditions
  \begin{align}
\tilde{Q}_\pm \equiv \frac12(1\pm i\gamma_0\gamma_5\tau^3) \quad  \begin{cases}
     \tilde{Q}_+\psi(x)\vert_{x_0=0} =0
     & \quad \tilde{Q}_-\psi(x)\vert_{x_0=T} =0\nonumber\\
   \bar \psi(x)\tilde{Q}_+\vert_{x_0=0} = 0   
    & \quad \bar \psi(x)\tilde{Q}_-\vert_{x_0=T} = 0
 \end{cases}
\end{align}
in time and  periodic ones in space. The massless fermion action is
\begin{equation*}
   S_f = a^4\sum_{x_0=0}^T\sum_{\bf x} \psibar(x)({\cal D}_W +\delta {\cal D}_W
)\psi(x)\,,
\end{equation*}
with ${\cal D}_W$ the standard Wilson fermion matrix and the boundary term
\begin{eqnarray*}
 \delta {\cal D}_W \psi (x)  &=& \left(\delta_{x_0,0}+\delta_{x_0,T}\right)
\Bigl[\left({\color{red} z_f}-1\right)+ \left(d_s-1\right) a{\bf
D}_s\Bigr]\psi(x)\,.
\end{eqnarray*}
The $\chi$SF boundary conditions can be derived from the standard SF boundary conditions by applying a non-anomalous chiral flavour rotation on the fermion doublet as follows:
  \begin{equation*}
R=
\left.\exp\left(i\dfrac{\alpha}{2}\gamma_5\tau^3\right)\right\vert_{\alpha=\pi/2}
\begin{cases}
  \psi &\to \psi^\prime= R\psi\\
  \psibar &\to\psibar^\prime= \psibar R\,.\end{cases}
\end{equation*}
Consequently composite operators, which depend on fermion fields, are also rotated: 
  \begin{equation*}
  O[\psi,\psibar] \to Q[\psi,\psibar]=
O[R\psi,\psibar R\,]    \,.
\end{equation*}
SF and $\chi$SF correlation functions of composite operators $O,Q$ defined in the bulk and  ${\cal O,Q}$
defined on a time boundary obey the following universality relation:
 \begin{eqnarray*}
  &\langle O \, {\cal O}\rangle_{\rm (SF)}^{\rm cont} &=\lim_{a \to 0} [Z_{O}
Z_{\cal O}\, \langle O \, 
  {\cal O}\rangle_{\rm (SF)} + {\rm O}(a)]\\
  &&=\lim_{a \to 0} [Z_{Q} Z_{\cal Q} \, \langle Q \, {\cal Q}\rangle_{(\chi
{\rm SF})} + {\rm O}({\color{red}a^2})]\, .
\end{eqnarray*}
Note that $\chi$SF incorporates automatic $\O(a)$ improvement.
The price to pay is the introduction of  boundary counterterms. We tune non-perturbatively the boundary 
counterterm coefficient  ${\color{red}z_f}$, while we fix the others at their tree-level value.

\subsection{Correlation functions}
As in ref.~\cite{DallaBrida:2018tpn}, we consider the set of fermion bilinear operators; e.g.
\begin{align*}
 V_{\mu}^{f_{1}f_{2}}(x)&=\overline{\psi}_{f_{1}}(x)\gamma_{\mu}\psi_{f_{2}}(x)
,
 &
A_{\mu}^{f_{1}f_{2}}(x)&=\overline{\psi}_{f_{1}}(x)\gamma_{\mu}\gamma_{5}\psi_{f_{2}}(x),
\end{align*}
with flavours $f_1,f_2 \in \{u,d,u',d'\}$, and determine the $\chi$SF
bulk-to-boundary correlation functions
\begin{align*}
  \gx^{f_1f_2}(x_0) &= -\frac12\left\langle X^{f_1f_2}(x){\cal
Q}_{5}^{f_2f_1}\right\rangle
\,, \qquad  \quad X=V_0,A_{0},S,P\,,\\
  \ly^{f_1f_2}(x_0) &= -\frac16\sum_{k=1}^3\left\langle Y_k^{f_1f_2}(x){\cal
Q}_{k}^{f_2f_1}\right\rangle\,, \qquad Y_{k}=V_{k},A_k,T_{k0},\widetilde{T}_{k0}\,.
\end{align*}
The complete list of boundary operators ${\cal
Q}_{5}^{f_2f_1}\,, {\cal
Q}_{k}^{f_2f_1}$ can be found in~
\cite{Brida:2016rmy}.

\noindent Up to discretization effects and boundary fields renormalization they
are related to the 
standard SF correlation functions $f_{\rm X}$ and $k_{\rm Y}$ by universality
\begin{alignat}{8}
 \fa^{\rm cont}\,  &=&\,  Z_{\rm A} \ga^{uu'} &=&\, Z_{\rm A} \ga^{dd'} &=&\,
-iZ_{\rm V}\gv^{ud} &=\, iZ_{\rm V}\gv^{du}, \\
 \fv^{\rm cont}\,  &=&\,  Z_{\rm V}\gv^{uu'} &=&\,   Z_{\rm V}\gv^{dd'} &=&\, -i
Z_{\rm A} \ga^{ud} &=\, i  Z_{\rm A}\ga^{du},\label{eq:fv}\\
\nonumber\\[0.02cm]
  \kv^{\rm cont}  &=&\,  Z_{\rm V}\lv^{uu'} &=&\,   Z_{\rm V}\lv^{dd'} &=&\, -i
Z_{\rm A}\la^{ud} &= \, i Z_{\rm A}\la^{du},\\
  \ka^{\rm cont}  &=&\,  Z_{\rm A} \la^{uu'} &=&\,   Z_{\rm A} \la^{dd'} &=&\,
-iZ_{\rm V}\lv^{ud} &= \, iZ_{\rm V}\lv^{du}\,.\label{eq:ka}
 \end{alignat}
The $\chi$SF correlation functions in eqs.~(\ref{eq:fv}),\,(\ref{eq:ka})  are ${\rm O}(a)$, since they become $\fv^{\rm cont},\ka^{\rm cont}$ in the continuum, which are parity odd.
The local  vector current can be replaced by the exactly conserved  one
${\widetilde V}_\mu(x)$ with normalization  $Z_{\rm \widetilde V}=1$.
Therefore $Z_{\rm A}$ may be obtained from the ratios
\begin{equation}
 Z^g_{\rm A} = {-ig_{\rm \widetilde V}^{ud}(x_0)\over
 \phantom{-i} g_{\rm A}^{uu'}(x_0)}\bigg|_{x_0=L/2}\quad
 \text{or}\quad
 Z^l_{\rm A} = {il_{\rm \widetilde V}^{uu'}(x_0)\over
 \phantom{i} l_{\rm A}^{ud}(x_0)}\bigg|_{x_0=L/2}\,.\label{eq:zaratios}
\end{equation}

\section{Computational setup and results}

We obtain results for $N_{\rm f}=3$ QCD in a non-unitary setup.
Valence quark propagators are inverted with $\chi$SF boundaries on the
configuration ensembles of~\cite{Campos:2018ahf},
generated on lattices with standard SF boundary conditions.
These configurations have been used for the RG-running of the quark mass
in a range of scales $2\,\mathrm{GeV} \lesssim \mu \lesssim 128\,\mathrm{GeV}$ ,
in the standard framework of finite-size scaling $L\to 2L$.

\subsection{Tuning}
We must ensure that massless QCD with $\chi$SF  boundary conditions is
correctly regularized. 
This is achieved by tuning the bare mass parameter $m_0$ to its critical
value, $m_{\rm cr}$, where the axial current is conserved, and by tuning
the boundary counterterm coefficient $z_f$ so that physical parity is restored.
At present we choose to set the PCAC mass to zero in terms of the SF correlation functions, so taking the $m_{\rm cr}$ value from~\cite{Campos:2018ahf},
and to set the $\chi$SF correlation function $\ga^{ud}$ to zero:
\begin{align}
m = {\tilde\partial_0 f_{\rm A}^{ud}(x_0) \over 
2f_{\rm P}^{ud}(x_0)}\bigg|_{x_0=L/2} &= 0, 
&\mcr \text{ tuning} \,,\nonumber \\
\color{red}\ga^{ud}(x_0)\bigg|_{x_0=L/2}  & \color{red}  = 0\,, & 
\color{red} \zf \text{ tuning}\,.\label{eq:zftuning}
\end{align}

By requiring that eq.~(\ref{eq:zftuning}) be satisfied, $z_f$ is tuned
for each ensemble as shown in Fig.~\ref{zf}.
\begin{center}
    \begin{tikzpicture}
    \node[anchor=south west,inner sep=0] (image) at (0,0)
{\includegraphics[width=0.65\linewidth]{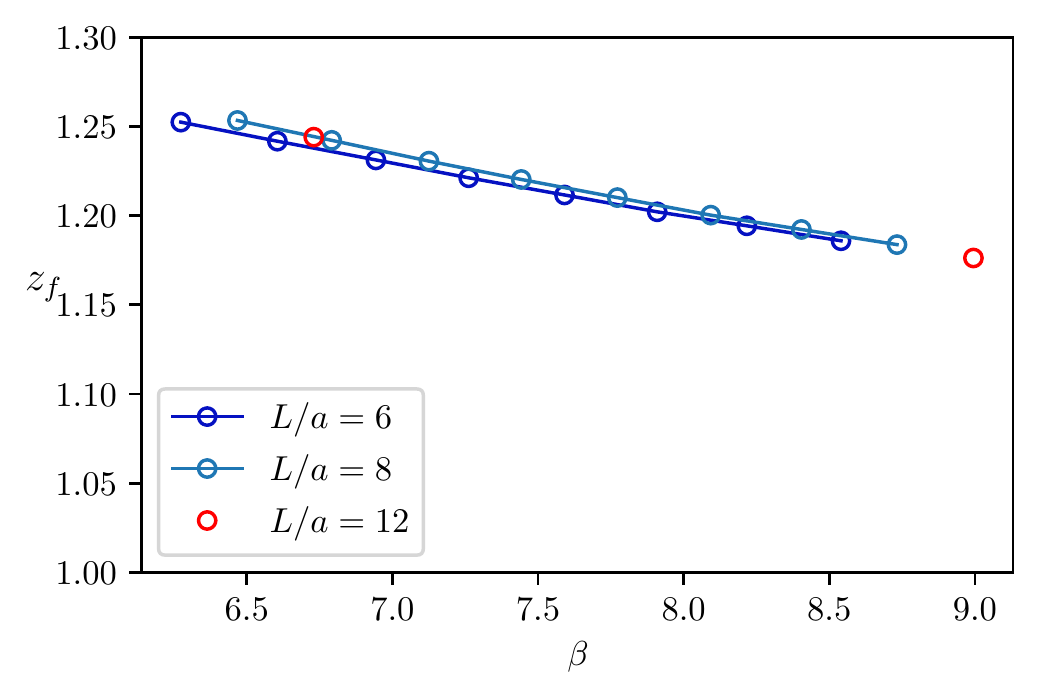}};
    \node[align=center,opacity=0.4,orange,rotate=30,font={\huge\bfseries}] at
(0.6\linewidth,0.20\linewidth) {\texttt{PRELIMINARY}};
    \end{tikzpicture}
    \vspace{-0.2cm}
\captionof{figure}{\label{zf}
Results of nonperturbative tuning of $z_f$,
according to eq.~\eqref{eq:zftuning}.}
\end{center}
\subsection{$\O(a)$-improvement}
Using eqs.~(\ref{eq:zaratios}) we obtain two estimates for $Z_{\rm A}$
\begin{equation*}
   Z_{\rm A}^g(\beta) = Z_{\rm A}^l(\beta) +  {\rm O}(a^2)
   \label{eq:g-vs-l}
\end{equation*}
which differ by discretization errors. In Fig.~\ref{zagl} we show the ratio of these two definitions and
we confirm that, after tuning $z_f$,
the ratio scales as $a^2$ and goes to 1 in the continuum.

\begin{center}
    \begin{tikzpicture}
    \node[anchor=south west,inner sep=0] (image) at (0,0)
{\includegraphics[width=0.65\linewidth]{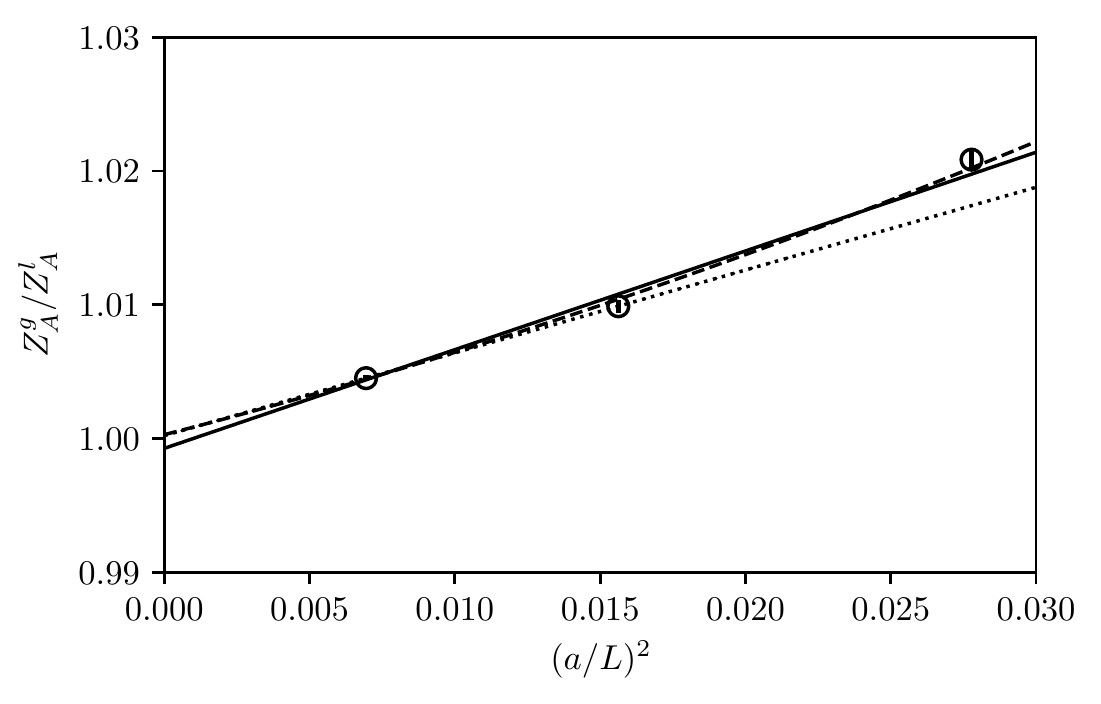}};
    \node[align=center,opacity=0.4,orange,rotate=30,font={\huge\bfseries}] at
(0.6\linewidth,0.20\linewidth) {\texttt{PRELIMINARY}};
    \end{tikzpicture}
    \vspace{-0.2cm}
    \captionof{figure}{\label{zagl} Ratio of two different
definitions of $Z_A$ (see eq.~\eqref{eq:zaratios})
calculated on our ensembles with $1/L =  4$ GeV.
}
\end{center}

Finally we study $Z^l_A(g_0^2)$ over the full range of ensembles. As seen in Fig.~\ref{za} a fit to the data matches
 onto the asympotic perturbative result in the limit $g_0^2 \to 0$.
\begin{center}
    \begin{tikzpicture}
    \node[anchor=south west,inner sep=0] (image) at (-6.0,0)
{\includegraphics[width=0.65\linewidth]{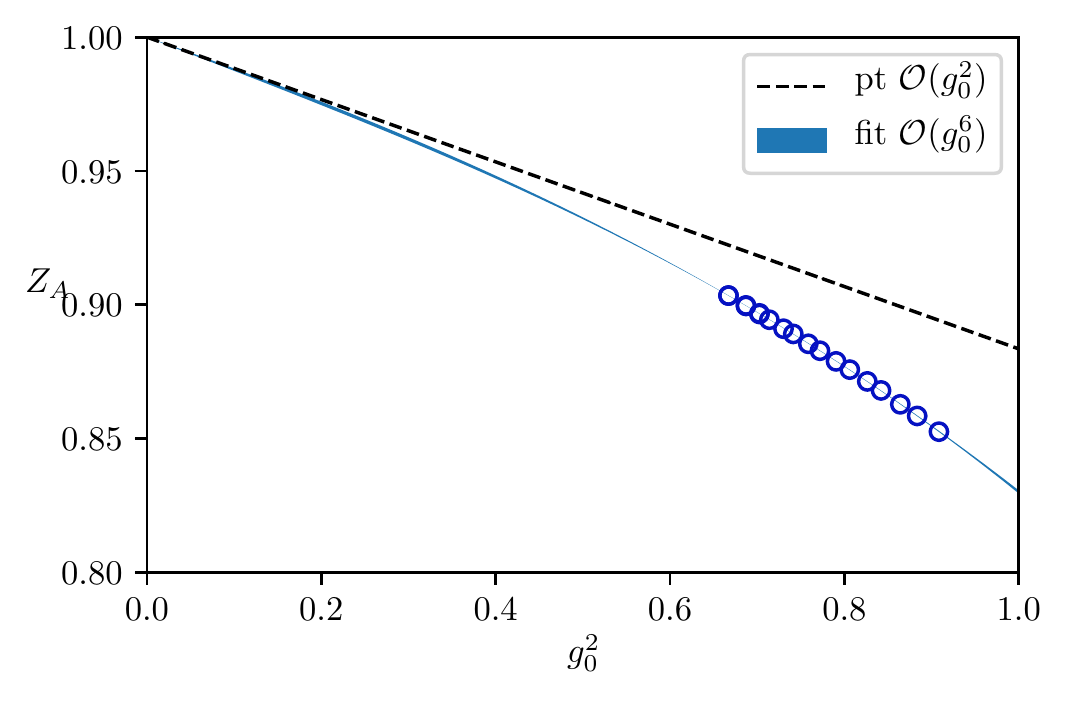}};
    \node[align=center,opacity=0.4,orange,rotate=-30,font={\huge\bfseries}] at
(-0.1\linewidth,0.20\linewidth) {\texttt{PRELIMINARY}};
    \end{tikzpicture}
 \captionof{figure}{\label{za} $Z^l_A(g_0^2)$ calculated across
 the full range of ensembles available. }
\end{center}
\section{Outlook for 4 fermions}
\subsection{Renormalization}
$\chi$SF framework is especially valuable in simplifying the
renormalization of four fermion operators.
They enter the most general  expression of the effective Hamiltonian which
describes flavour physics processes at low energy
in the Standard Model (SM) and its extensions (BSM). Here we focus on  $\Delta
F=2$ transitions.
The 4-quark operators
with four distinct flavours
\begin{equation*}
O_{XY}^{\pm} \equiv {1\over2}\left[
(\overline\psi_{1}\Gamma_{X}\psi_{2})(\overline\psi_{3}\Gamma_{Y}\psi_{4})
\pm (2\leftrightarrow4)\right]
\end{equation*}
can be classified as parity even and parity odd:
\begin{align*}
O_{k}^{e,\pm} &\in \left\{ O_{VV+AA}^{\pm}, O_{VV-AA}^{\pm},
O_{SS-PP}^{\pm}, O_{SS+PP}^{\pm},
O_{T T}^{\pm}\right\},\nonumber\\
{O}_{k}^{o,\pm} &\in \left\{ {O}_{VA+AV}^{\pm},{O}_{VA-AV}^{\pm},
{O}_{SP-PS}^{\pm}, {O}_{SP+PS}^{\pm},
{O}_{T\tilde T}^{\pm}\right\}.
\end{align*}
Due to the explicit breaking of chiral symmetry of the Wilson regularisation,
the operators in ge\-ne\-ral mix as follows:
\begin{align*}
O_{i}^{e,\pm} &= 
\sum_{jm}\, Z_{ij}^{e,\pm} \,(\delta_{jm}+\Delta_{jm}^{e,\pm} )\,
O_m^{e,\pm}\,,\\
O_{i}^{o,\pm} &= 
\sum_{jm}\, Z_{ij}^{o,\pm}
\,(\delta_{jm}+\ccancel[red]{\color{red}\Delta_{jm}^{o,\pm}}
)\,  O_m^{o,\pm}\,.
\end{align*}
The parity-odd sector has a simpler, continuum-like mixing pattern
($\Delta_{jm}^{o,\pm} =0$)~\cite{Donini:1999sf}.
In previous Wilson fermion computations~\cite{Guagnelli:2005zc,Papinutto:2016xpq}, standard SF renormalization conditions were imposed 
on parity-odd operators
by setting  suitable renormalized correlation functions  equal
to their tree level values at the scale $\mu=L^{-1}$: 
 \begin{align}
 \label{eq:4ptF}
F_{i}(x_{0})=\langle \mathcal{O}_{5}^{\prime 4 5}  {
O}_{i}^{o,1234}(x_{0})\mathcal{O}_{5}^{2 1} \mathcal{O}_{5}^{53} \rangle \,.
\end{align}
Note that five distinct valence flavours are required; see the left diagram in Fig.~\ref{fig:diagrams}. 
These four-point correlation functions are parity even, but suffer from large statistical fluctuations. 
Moreover, they have bulk $\O(a)$ discretization errors.

Alternatively we plan to employ a new renormalization scheme on $\chi$SF three-point correlation functions
 \begin{align}
\label{eq:4ptG}
G_{i}(x_{0})=\langle \mathcal{Q}_{5}^{\prime 2 1} {
Q}_{i}^{o,1234}(x_{0})\mathcal{Q}_{5}^{43} \rangle \,,
\end{align}
which are statistically less noisy and automatically $\O(a)$ improved in the bulk.
 Performing suitable chiral rotations, the $\chi$SF flavours are
rotated
into the physical flavours~\cite{Mainar:2016uwb}. Thus we can map the renormalized
$[G_{i}(x_{0})]_{\rm R}$ into
the continuum correlation function
\begin{align*}
[G_{i}(x_{0})]_{\rm R} \rightarrow  \langle
\mathcal{O}_{5}^{'21}O_{i}^{e,1234}(x_{0})\mathcal{O}_{5}^{43} \rangle^{\rm
cont} .
\end{align*}

\begin{center}
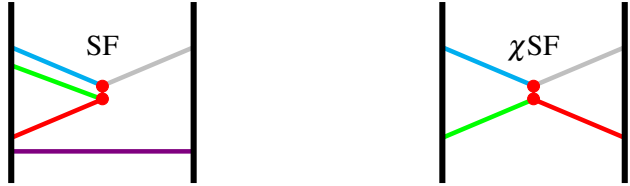

\begin{tikzpicture}[scale=0.6]

\draw[thick,green,line width=0.6mm] (1.0,3.6) -- (3.0,2.85);

\draw[thick,red,line width=0.6mm] (1.0,2.0) -- (3.0,2.85);
\draw[thick,violet,line width=0.6mm] (1.0,1.7) -- (5.0,1.7);

\draw[thick,cyan,line width=0.6mm] (1.0,4.0) -- (3.0,3.15);
\draw[thick,lightgray,line width=0.6mm] (5.0,4.0) -- (3.0,3.15);
\filldraw [red] (3.0,2.86) circle [radius=1.3mm] (3.0,3.141) circle
[radius=1.3mm];
\node[anchor=north, black] at ({3.0},{4.5}){SF};
\node[anchor=north, black] at ({3.0},{3.0}){$$};
\draw[draw=black,line width=0.8mm] (1.0,1.0) --  (1.0,5.0);
\draw[draw=black,line width=0.8mm] (5.0,1.0) --  (5.0,5.0);
\end{tikzpicture}
\hspace{3.cm}
\begin{tikzpicture}[scale=0.6]
\draw[thick,green,line width=0.6mm] (1.0,2.0) -- (3.0,2.85);
\draw[thick,cyan,line width=0.6mm] (1.0,4.0) -- (3.0,3.15);
\draw[thick,red,line width=0.6mm] (5.0,2.0) -- (3.0,2.85);
\draw[thick,lightgray,line width=0.6mm] (5.0,4.0) -- (3.0,3.15);
\filldraw [red] (3.0,2.86) circle [radius=1.3mm] (3.0,3.141) circle
[radius=1.3mm];
\node[anchor=north, black] at ({3.0},{4.5}){$\chi$SF};
\draw[draw=black,line width=0.8mm] (1.0,1.0) --  (1.0,5.0);
\draw[draw=black,line width=0.8mm] (5.0,1.0) --  (5.0,5.0);
\end{tikzpicture}
\captionof{figure}{\label{fig:diagrams}
Correlation functions used in two renormalization schemes. Left: parity odd operators with standard SF boundary conditions; right: parity odd operators with $\chi$SF boundary conditions. Different colors stand for different valence flavors.}\end{center}

\noindent
\subsection{Physical determinations}
The renormalization program will be employed in the lattice computation of the
physical $B_{\rm K}$-parameter which controls
the $\bar K^0-K^0$ meson oscillations, towards a better understanding of the
physics of CP violation in the SM and BSM. Whereas there is general agreement among various collaborations
on $B_{\rm K}$ in the SM, the situation is somewhat unclear for the BSM contributions~\cite{Aoki:2019cca}.  

In the SF approach, $B_{\rm K}$ has been computed 
by combining the renormalization parameters based on eq.~(\ref{eq:4ptF})
with bare four-point correlation functions of the  parity odd
operators in a twisted mass QCD setup~\cite{Dimopoulos:2009es,Frezzotti:2004wz}.

We plan to use the $\chi$SF renormalization conditions based on eq.~(\ref{eq:4ptG}). For the bare matrix elements
we will employ the CLS ${N_{\rm f}=2+1}$ ensembles,
  characterised by {\it large} physical 
volumes with open boundary conditions and by {\it non-zero} quark masses~\cite{Bruno:2014jqa,Bruno:2016plf}.
The sea quarks are \mbox{Wilson}/Clover. Valence fermions are fully 
twisted~\cite{Bussone:2018ljj}, with three flavours tuned at twisted angle
$\alpha=\pi/2$ and the fourth one at $\alpha=-\pi/2$.
Unitarity is lost at finite lattice spacing, but it is recovered in the 
continuum limit. 
Performing distinct {\it Osterwalder-Seiler} chiral rotations for each flavour, 
correlation functions with parity-odd operators
(renormalized in the $\chi$SF scheme as outlined above) are mapped onto 
the 3-point correlation functions
of parity even operators with pseudoscalar sources, 
from which the $B$-parameters are readily extracted:
\begin{align*}
B_{{\rm K}i}(\mu) \propto \langle \bar K^0 | \; [O^e_i (\mu)]_R\; |  K^0\rangle
\,.
\end{align*}
\section{Acknowledgements}
We gratefully acknowledge help from Patrick Fritzsch, Carlos Pena, David Preti, and \mbox{Alberto} Ramos.
AL would like to thank the conference organisers, and F. Joswig, S. 
Kuberski, and R. Sommer for useful discussions.
We acknowledge the Santander Supercomputacion support group at the University of
Cantabria which provided access to the Altamira Supercomputer at the
Institute of Physics of Cantabria (IFCA-CSIC), member of the Spanish
Supercomputing Network, for performing simulations/analyses. Special thanks go to Aida Palacio Hoz
for her constant help.

\end{document}